\documentclass[12pt]{article}

\usepackage{sbc-template}
\usepackage{graphicx,url}
\usepackage{xcolor,float}
\usepackage{float}   
\usepackage{comment}

\usepackage{hyperref}
\title{A Low-cost IoT Architecture to support Urban Mobility for Visually Impaired People}

\author{Nádia Aparecida de Oliveira Silva\inst{1}, Rodrigo Moreira\inst{1,2}, \\ Larissa Ferreira Rodrigues\inst{2}, Rafael Marinho e Silva\inst{2}}

\address{Institute of Exact and Technological Sciences (IEP) \\Federal University of Viçosa (UFV), Rio Paranaíba, MG, Brazil
\nextinstitute
Faculty of Computing (FACOM) \\ Federal University of Uberlândia (UFU), Uberlândia, MG, Brazil
  \email{\{nadia.silva,rodrigo\}@ufv.br,\{larissarodrigues, rafael.marinho\}@ufu.br}}

\begin{document} 

\maketitle

\begin{abstract}

People with visual impairments struggle with urban mobility and independent travel, opening up opportunities for technological advances to improve their quality of life. The Internet of Things (IoT) plays an essential role in bringing improvements and accessibility for visually impaired people. Although alternatives aimed to use IoT in urban mobility, those solutions are still in the initial stages and do not supports urban mobility for people with visual impairment. This paper proposed and evaluated a low-cost IoT architecture that uses Single-Border Computers (SBCs) to support urban mobility. A performance evaluation showcased that our low-cost architecture handles bus trace workload and is suitable for supporting impaired people to get information concerning bus location on Smart Cities scenarios.

\end{abstract}
    
\section{Introduction}

It is estimated that approximately 2.2 billion people worldwide are affected by visual impairment. In almost half of the cases, at least 1 billion visual impairment could have been prevented or treated early \cite{WHO2021}. People with impairments face difficulties with urban mobility and independent travel, opening up opportunities to enhance the quality of life and social inclusion. In Brazil, urban mobility occurs predominantly by urban and metropolitan buses. According to Institute for Applied Economic Research (Ipea), 65\% of Brazilians who live in capital cities use public transportation to perform their daily activities \cite{Galindo2019}.


In this context, the Internet of Things (IoT) plays an essential role in supporting improvements and accessibility for the visually impaired people, raising as an alternative with lower investments and attractive for developing countries \cite{Vargas2020}. Although alternatives have been proposed to use IoT in urban mobility, those solutions still in initial stages, and guidelines have not yet been developed to guarantee mobility for people with visual impairment \cite{Baucas2021}.


This paper proposed a low-cost IoT architecture that uses single-border computers (SBCs) to enable intelligence at the edge. Although existing long-range communication technologies such as Long Term Evolution (LTE), our architecture relies on Long-Range (LoRa)\footnote{\url{https://lora-alliance.org/}}. LoRa strengthens our architecture, supporting a communication energy-aware covering long distances. Our IoT architecture allows visually impaired users to be notified audibly and mechanically about a bus's proximity. Unlike the works found in the literature, which predominantly deal with individual geospatial positions, we go beyond with an architecture that allows disabled and non-disabled users to follow the busses route on the map.

Furthermore, a novel contribution of this paper is an IoT architecture assessment considering low-cost devices as a technological enabler for improving urban mobility. We evaluated the suitability of the low-cost device to handle simulated workloads similar to real ones. Our results appoint that the proposed architecture is suitable for handling geographic location data at the edge and enabling applications for Smart Cities. 

The remainder of this paper is organized as follows. Section \ref{sec:related_work} presents related work. Section \ref{sec:proposal_architecture} describes our proposed architecture. The experimental evaluation is presented in Section \ref{sec:experimental_eval}. Section \ref{sec:results} presents and discusses the results. Finally, we provide concluding remarks and future work in Section \ref{sec:concluding}.

\section{Related Work} \label{sec:related_work}


\cite{Kim2016} proposed a navigation system based on Bluetooth low energy (BLE) that provides turn-by-turn voice directions inside Tokyo Station to help blind users. However, this approach is limited by low-range technologies. We overcome this limitation using LoRa that reaches longer distances becoming the proposed architecture scalable.

\cite{Gayathri2018} developed a mobile application that integrated an IoT system capable of tracking buses, using LoRa, GPS, cloud computing, and Arduino. This work is similar to our proposal because it aims to track a vehicle in real-time and uses related technologies. However, the application is not accessible for the visually impaired and does not consider strategies to deal with latency between a cloud server and Arduino.

\cite{Andrade2019} proposed a bus detection system based on BLE to improve the traveling of blind people in large urban centers. The proposed system comprises hardware embedded in the buses and a mobile application for users with visual impairment. Due to the distance limitations present in the BLE technology, the authors used Decision Tree and K-NN classifiers to identify the approach and arrival of the bus. In contrast, our approach considers a GPS-based location ensuring a more accurate location, not subject to probabilistic bias that can occur in strategies that use supervised classifiers.

\cite{Murugeshwari2020} proposed a method to identify bus names and informing the visually impaired by voice signal the name of the approaching bus line. The Radio Frequency (RF) Transceiver was putting in each bus stop, and another RF receiver was setting on the cane to inform the visually impaired when the bus arrives. In contrast, our method presents the route along the way.  This feature allows the architecture to be used by people with and without disabilities. 

\cite{Salazar2021} developed an integrated framework with an IoT architecture customized for an electronic cane project created in Brazil to help visually impaired people in terms of orientation and mobility in smart cities. However, they exploited only spatial information from the environment, improving only the mobility of visually impaired people. On the other hand, our study improving urban mobility access.

\section{Our Proposal Architecture} \label{sec:proposal_architecture}

Aiming to improve urban mobility for visually impaired people, we propose an architecture based on low-cost IoT devices. This architecture can be implemented in scenarios of intelligent cities and has a technological opening to compose other application ecosystems. An open IoT environment enables vertical applications such as e-Gov and intelligent constructs to be exploited based on sensor data placed in specific contexts.

\begin{figure}[!htbp]
\begin{center}
\includegraphics[width=0.45\linewidth]{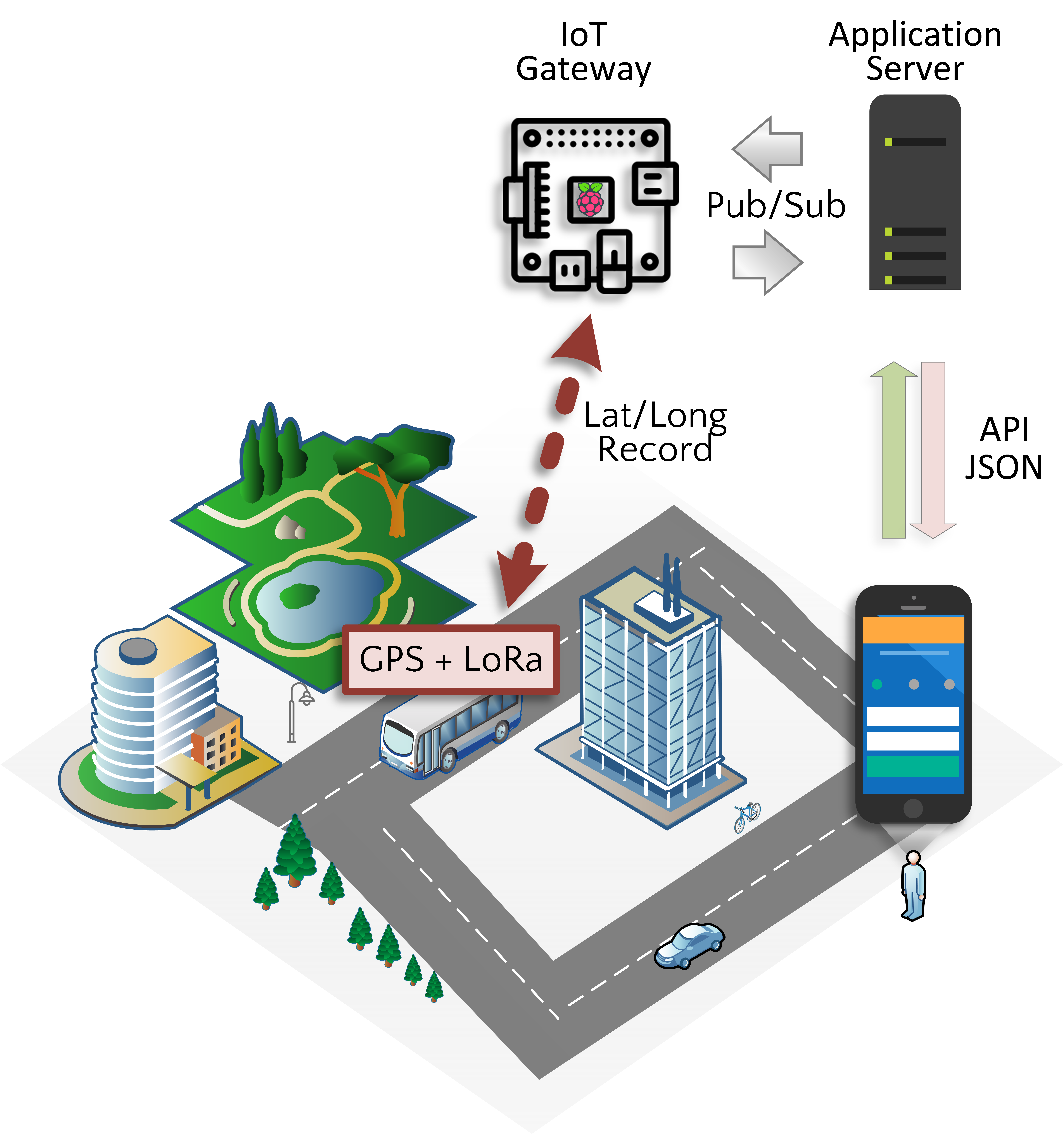}
\caption{Proposed Architecture Scenario.}
\label{fig:scenario}
\end{center}
\end{figure}

In Figure~\ref{fig:scenario} we represent the scenario of an architecture to improve urban mobility for people with visual impairments. The architecture consists of three fundamental components: a GPS NEO-6M location sensor, ESP32 (ESP-WROOM-32) LoRa, and power supply embedded in the transport system buses, a gateway IoT based on single board computing (Raspberry Pi3), and an application server IoT running on a cloud computing system. 

As a public transport buses move, it sends information regarding its trace to the gateway IoT. Each packet contains data such as latitude and longitude sent regularly in each five (5) seconds. The gateway preprocesses the location information and publishes it to the IoT Urban Mobility application that runs in the cloud. On the client-side, a mobile application containing an IoT application that signs the context of the urban mobility application notifies the user with audible and mechanical signals of the proximity of the public transport vehicle.

\section{Experimental Evaluation Method} \label{sec:experimental_eval}

\begin{figure}[!htbp]
\begin{center}
\includegraphics[width=0.52\linewidth]{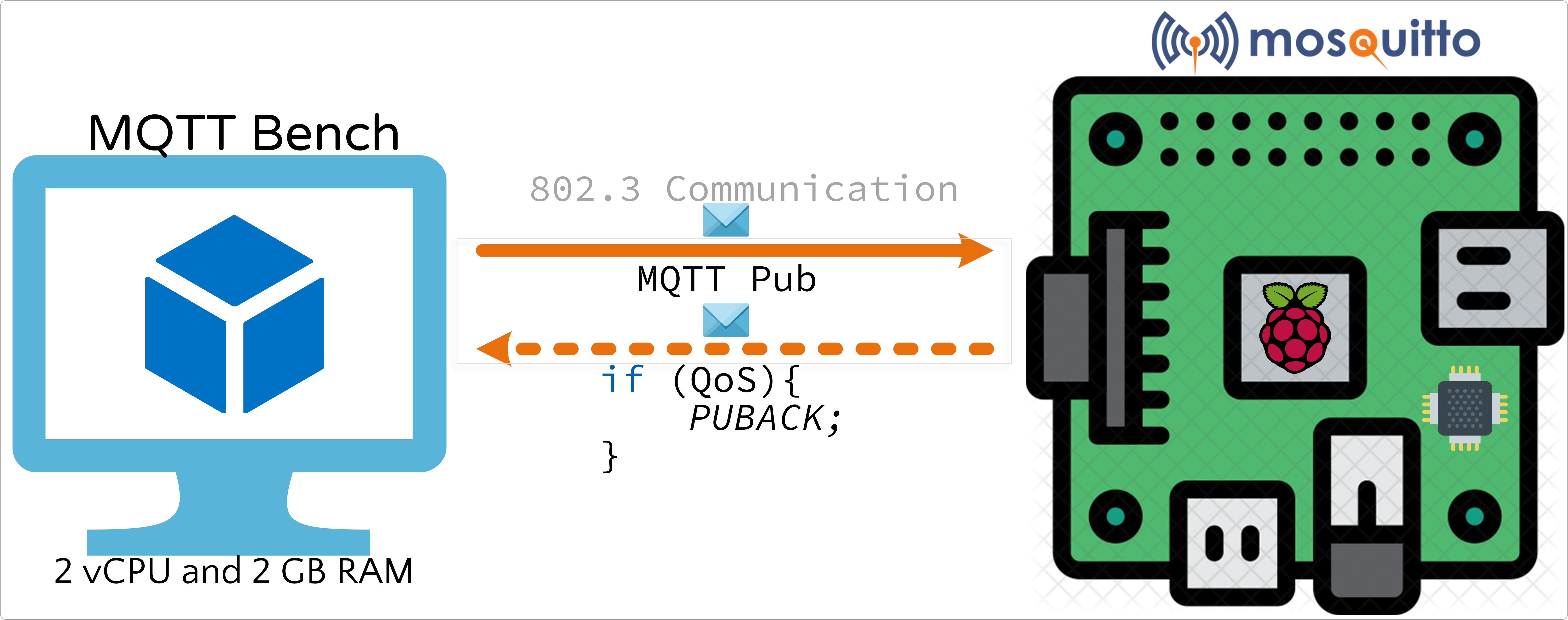}
\caption{Experimental Testbed.}
\label{fig:experimental_testbed}
\end{center}
\end{figure}

To validate our architecture, we carried experiments to measure architecture performance to deal with workloads inherent in the context of public transport. In Figure~\ref{fig:experimental_testbed} we present our testbed topology and entities, which are directly connected by ethernet technology.

A workload was directed to the IoT gateway based on Mosquitto~\cite{Light2017}, running on a Raspberry Pi3 with Ubuntu 20.04 LTS operating system. We measure the proposed architecture capability in request responding for message publications in the Message Queuing Telemetry Transport (MQTT) pattern using the MQTT tool Bench\footnote{Available at \url{https://github.com/krylovsk/mqtt-benchmark }.}. This benchmark tool ran on an Ubuntu 20.04 LTS virtual machine with 2vCPU flavor and 2GB RAM.

Measuring the IoT gateway performance is relevant because it has limited resources compared to a server running in a remote cloud. The gateway is responsible for consolidating information from all urban transport vehicles and has limited resources. At this point, it is necessary to assume that the workload may be high since, in urban transport, there are numerous lines and vehicles where the current trace and distance may be relevant to users. 

We carried experiments considering that messages are transmitted with QoS and without QoS. In the MQTT protocol, QoS is divided into three levels: 0 - at most once (there is no guarantee of delivery), 1 - at least once (guarantees that a message is delivered at least one time), and 2 - exactly once (guarantees that each message is received only once by the intended recipients). In this paper, we assess levels 0 and 2. 

We ran experiments to answer the following questions: 1) Is the increase in the latency of MQTT messages related to the message size sent by the buses? 2) Are the messages number sent by the bus directly related to the latency and CPU utilization of low-cost IoT devices? 3) What processing overhead does the addition of Quality of Service (QoS) impose on Low-Cost devices?

\section{Results and Discussion} \label{sec:results}

The first experiment measured the relationship between message size (\textit{bytes}) and the latency an MQTT message experiences. With the \textit{Pearson} correlation, considering the average latencies of 10 experiments, we found that there is a strong correlation since the measured coefficient is $0.999$. Therefore, it is possible to admit that latency increases as the size (\textit{bytes}) of messages sent by buses increases.

To check the relationship between the number of messages sent and the latency and CPU utilization of the architecture's IoT gateway. The measured results suggest a strong correlation between the message number and the IoT gateway CPU consumption since the measured coefficient was $0.729$. 

\begin{figure}[!htbp]
\begin{center}
\includegraphics[width=0.45\linewidth]{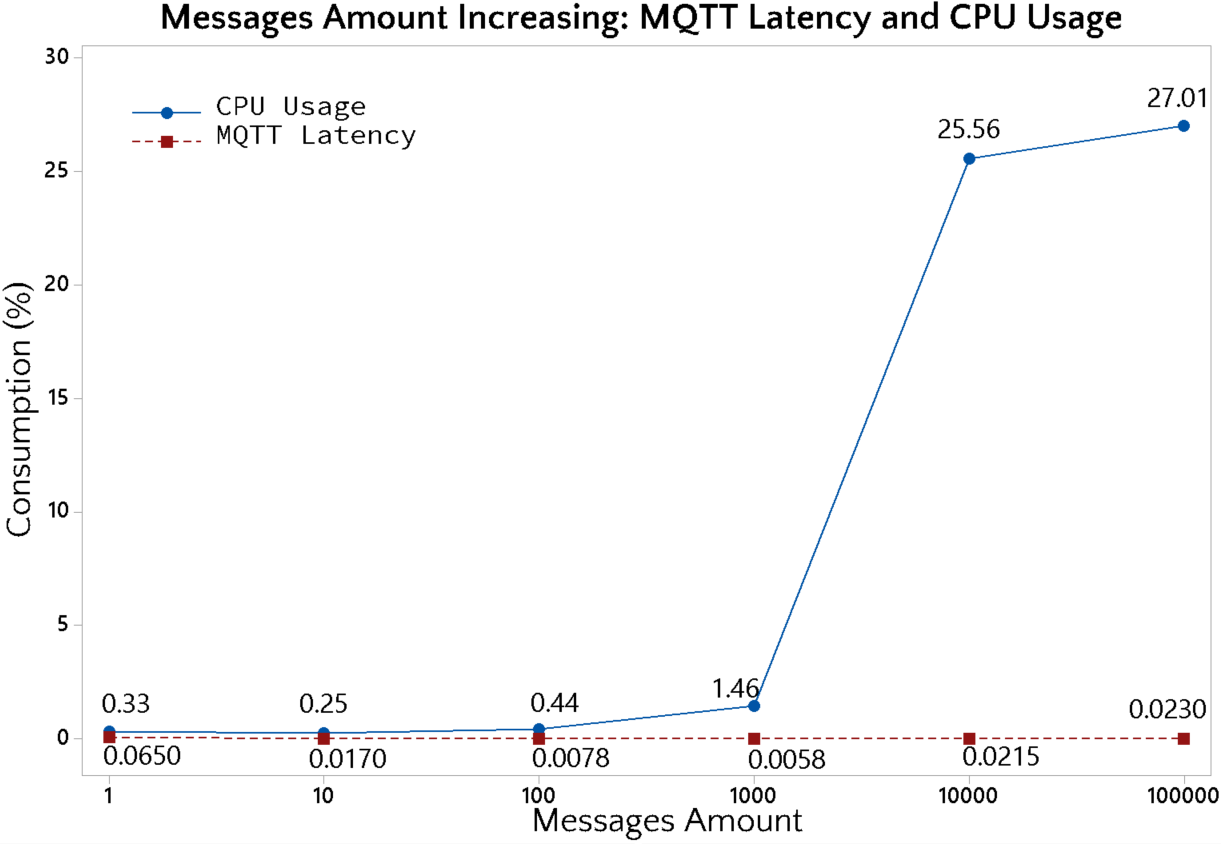}
\caption{Messages Amount Increasing: CPU Consumption vs. MQTT Latency.}
\label{fig:graph_2-latency-cpu}
\end{center}
\end{figure}

The exact relationship does not exist between the number of messages and the latency that MQTT messages experienced. The measured coefficient is $-0.017$, representing a weak negative relationship, as in Figure~\ref{fig:graph_2-latency-cpu} that as the message number increased, only the CPU consumption responded positively. In this observed scenario, we attribute the increase in CPU consumption to the number of interruptions that the message's arrival causes.

We evaluated the processing overhead when imposing QoS on the messages transmitted by the buses. As Figure~\ref{fig:graph_1}, the average consumption to process messages from 1,000 buses without QoS was $10.47\%$ whereas when QoS was imposed for MQTT messages of type Level 2, the CPU consumption was $23.94\%$. 

With the confidence interval of $95\%$, it is possible to assume that there is indeed an increase in CPU consumption applying QoS. This QoS demand caused an average increase of $\approx2.2 \ times$ when considering 1,000 buses. Alternatively, we evaluate a scenario containing a smaller amount of buses. Finally, we measure the impact of applying QoS into MQTT messages in a scenario of few buses, 100. According to Figure~\ref{fig:graph_3}, we observed that there was no statistically higher overhead when transmitting messages with QoS—leading us to assume that the addition of QoS only becomes significant when the IoT gateway deals with large amounts of the bus.

\begin{minipage}{0.9\textwidth}
    \centering
    \begin{minipage}{0.45\linewidth}
        \begin{figure}[H]
            \includegraphics[width=0.8\linewidth]{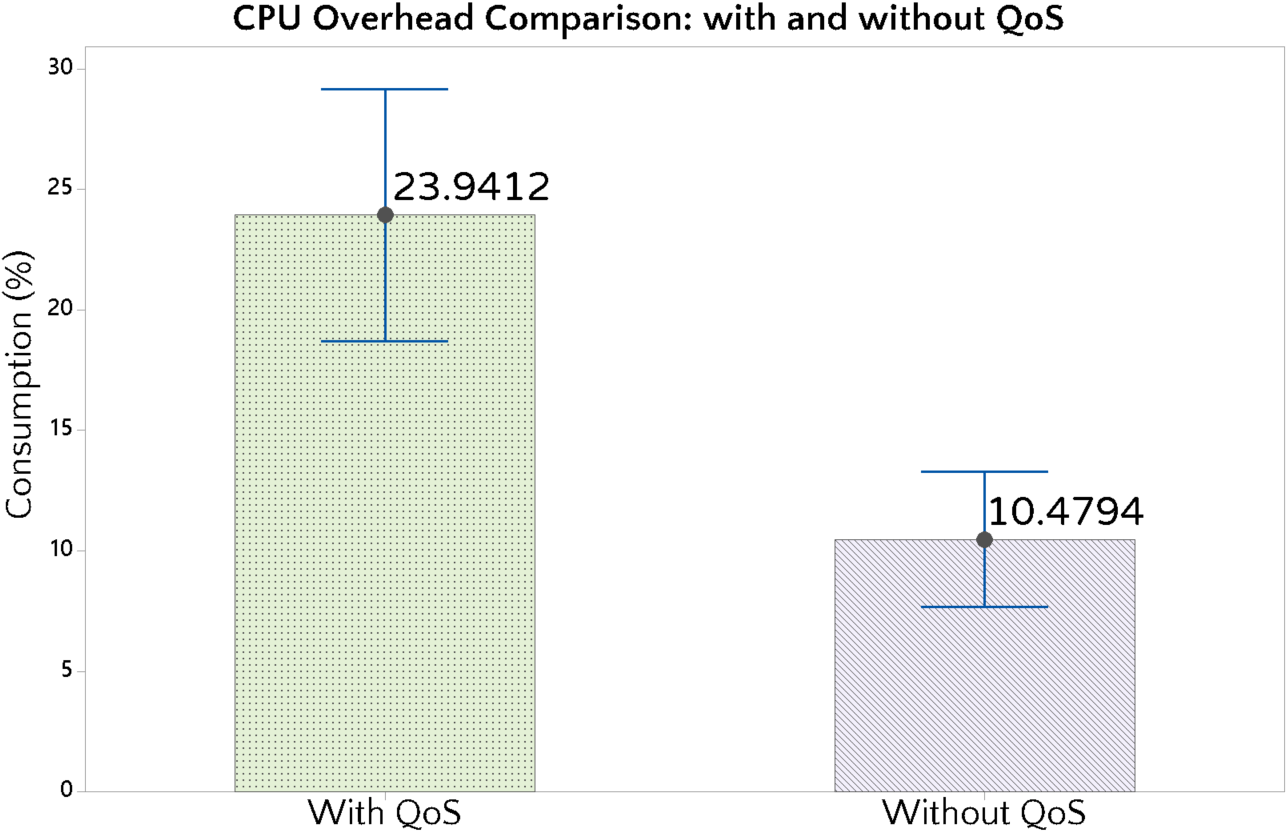}
            \caption{CPU Overhead Comparison.}
            \label{fig:graph_1}
        \end{figure}
    \end{minipage}
    \hspace{0.03\linewidth}
    \begin{minipage}{0.5\linewidth}
        \begin{figure}[H]
        \includegraphics[width=0.74\linewidth]{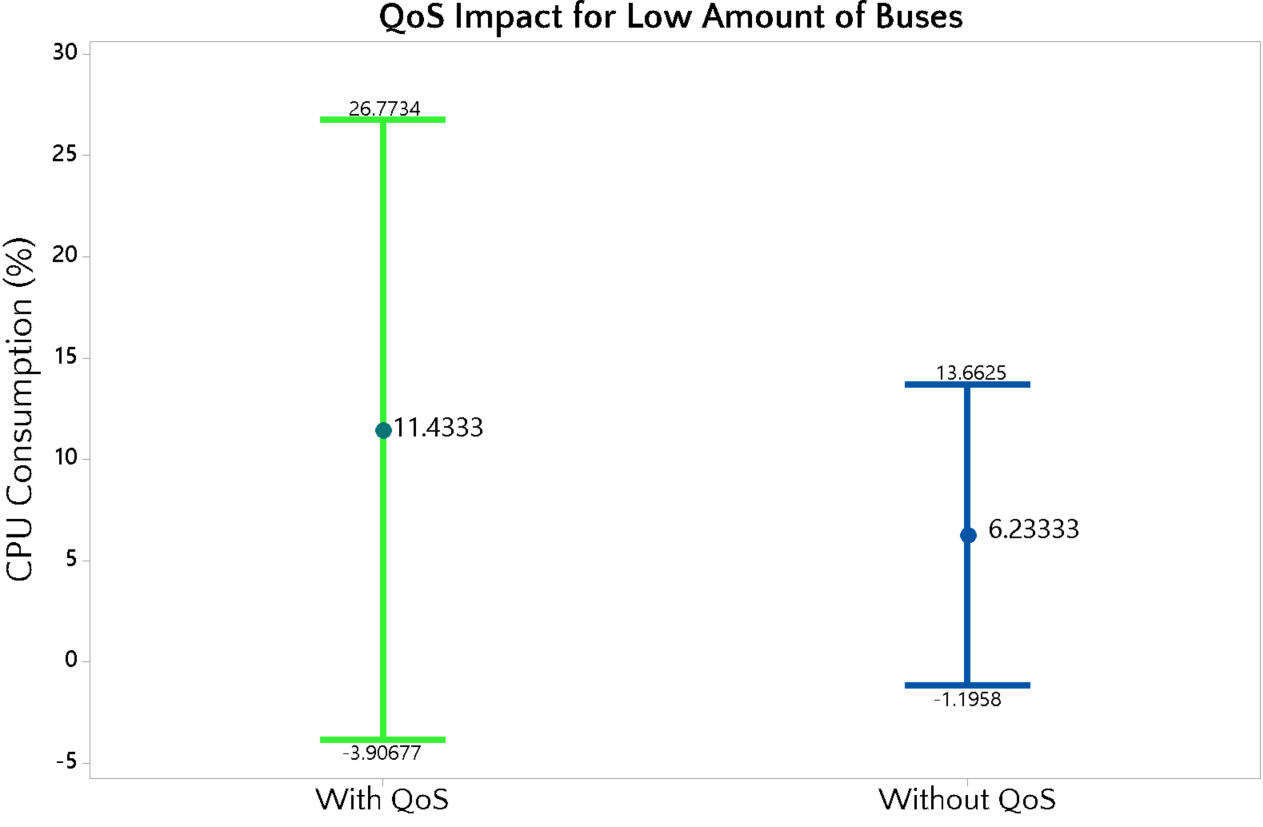}
        \caption{Impact of QoS in Low Amount of Buses.}
        \label{fig:graph_3}
    \end{figure}
\end{minipage}
\end{minipage}

\section{Concluding Remarks} \label{sec:concluding}

In this paper, we proposed and evaluated a three-tier low-cost IoT architecture. Carried experiments allowed notice that message increasing does not degrade the reliability of incoming messages to the IoT gateway. Furthermore, we point out a non-linear relationship between the number of buses and the processing time with and without QoS. In scenarios with few buses (100), QoS's overhead imposes on the CPU is insignificant. Consequently, IoT gateways are sufficiently capable of handling the workloads that embedded sensors on buses impose. On the other hand, as the number of buses increases (1,000), more CPU consumption occurs to transmit messages with QoS.

As future work, we need to evaluate other mechanisms for data transmission over long distances, an alternative to LoRa, such as mobile networks. It is also essential to assess the differences between the IoT gateways available in the literature, looking for availability and reliability concerns. We also consider evaluating the suitability of applying artificial intelligence to combine traffic-related information to predict delays and estimated arrival times.

\section*{Acknowledgments}
This study was financed in part by the Coordenação de Aperfeiçoamento de Pessoal de Nível Superior – Brasil (CAPES) – Finance Code 001.

\bibliographystyle{sbc}
\bibliography{sbc-template}

\end{document}